\documentclass[twocolumn,showpacs]{revtex4}
\usepackage{graphics}
\usepackage{amsmath}
\usepackage{amsfonts}
\usepackage{amssymb}
\usepackage{epsf}
\usepackage{hyperref}

\begin{document}
\title{Evaporation of an atomic beam on a material surface}
\author{G. Reinaudi, T. Lahaye, A. Couvert, Z. Wang, and D. Gu\'ery-Odelin}
\affiliation{Laboratoire Kastler Brossel$^{*}$, 24 rue Lhomond,
F-75231 Paris Cedex 05, France}
 \date{\today}

\begin{abstract}
We report on the implementation of evaporative cooling of a
magnetically guided beam by adsorption on a ceramic surface. We
use a transverse magnetic field to shift locally the beam towards
the surface, where atoms are selectively evaporated. With a 5~mm
long ceramic piece, we gain a factor $1.5\pm0.2$ on the phase
space density. Our results are consistent with a 100\% efficiency
of this evaporation process. The flexible implementation that we
have demonstrated, combined with the very local action of the
evaporation zone, makes this method particularly suited for the
evaporative cooling of a beam.
\end{abstract}

\pacs{32.80.Pj, 03.75.Pp}

\maketitle

Since its first demonstration for magnetically trapped atomic
hydrogen \cite{hydrogen}, evaporative cooling has proven to be a
powerful technique to increase the phase-space density of trapped
gases. It was used, in the case of alkali vapors, to reach the
Bose-Einstein condensation (BEC) threshold \cite{rmpnobelbec}.
This cooling technique plays a central role in the rapidly
expanding field of ultracold quantum degenerate gases. Evaporative
cooling occurs when energetic atoms are removed from the cloud as
a result of elastic collisions. Since these atoms belong to the
high energy tail of the thermal distribution, the remaining
trapped atoms collisionally equilibrate to a lower temperature.

As initially proposed and studied theoretically in Ref.
\cite{Mandonnet00}, if a non-degenerate, but already slow and cold
beam of particles, is injected into a magnetic guide where
transverse evaporation takes place, quantum degeneracy can be
achieved at the exit of the guide. Such a scheme transposes in the
space domain what is usually done for trapped atoms in time, so
that all operations leading to condensation are performed in
parallel, resulting in a much larger expected output flux.

In recent experiments, evaporative cooling of a beam has been
implemented by driving transitions to an untrapped state with
radio-frequency \cite{prlrb2} or/and microwave \cite{lahaye05}
fields. The drawback of these two methods lies in the range over
which a radio-frequency antenna or a microwave horn effectively
acts. On our experimental setup, atoms are affected by the field
on a zone whose length is at least 20 cm. Such a range limits the
number of evaporation zones that can be practically implemented.
In addition, it turns out that it is quite difficult to ensure a
100\% efficiency of the evaporation with those two methods, which
strongly affects the possible gains on the phase space density and
on the collision rate.

In this article we report on a much more local evaporation
technique -- the evaporation zone is 5~mm long. It relies on the
elimination of atoms on a surface. This technique has been used in
Ref. \cite{cornell} to evaporatively cool atoms to Bose Einstein
Condensation with a dielectric surface. We demonstrate here a
possible implementation of this technique to a magnetically guided
atomic beam, and discuss its prospects.

The experimental setup, described in detail in Refs.
\cite{prlrb2,lahaye05}, allows for the generation of an intense
and continuous beam of cold $^{87}$Rb atoms polarized in the $|F =
1, m_F = -1\rangle$ state, and magnetically guided over 4.5 m.
This beam results from the overlapping of packets of cold atoms
injected into the magnetic guide at a rate of 5 per second. Each
packet is prepared by loading an elongated magneto-optical trap
(MOT) that collects $2\times 10^9$ atoms from a Zeeman slower
\cite{zeemanslower} in 100~ms. The packets are launched at a
velocity of 1.1 m.s$^{-1}$, towards the magnetic guide entrance,
by a moving molasses technique \cite{launch}.

\begin{figure}[t!]
\centering\begin{center} \mbox{\epsfxsize 3.4 in
\epsfbox{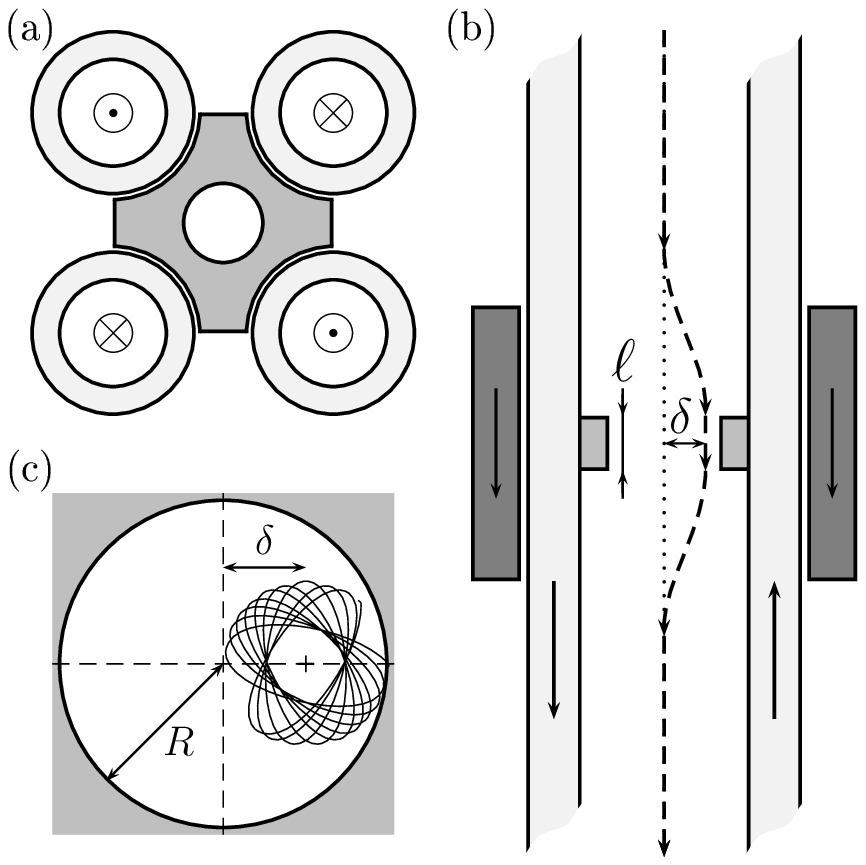} } \caption{(a) Section cut of the four copper
tubes that generate the two-dimensional quadrupole of the magnetic
guide, with the cross-shaped ceramic piece. (b) Deflection of the
minimum of the magnetic confinement using a pair of coils. (c)
Typical atomic trajectory that precesses around the minimum of the
magnetic potential because of the non-harmonic
confinement.}\label{figure1}
\end{center}
\end{figure}

The guide, placed inside the vacuum system, consists of four
parallel, water-cooled copper tubes, which are held together using
5~mm long ceramic pieces every 40 cm. The cross shape of the
ceramic piece (see Fig. \ref{figure1}.a) permits to accommodate
the four copper tube (outer diameter 6 mm, inner diameter 4~mm).
Atoms propagate through a cylindrical hole of radius $R=1.5$~mm.
The currents run are chosen with opposite sign for adjacent tubes
in order to generate a two-dimensional quadrupole magnetic field
configuration characterized by a transverse gradient $b=800$~G/cm,
with a current of 320~A per tube. In addition we superimpose a
longitudinal bias field $B_0=1$~G to avoid Majorana spin-flips
losses \cite{lahaye05}. The resulting semi-linear confining
potential reads:
\begin{equation}
U(r)=\mu\sqrt{B_0^2+b^2r^2}, \label{pot}
\end{equation}
where $\mu=\mu_B/2$ is the magnetic moment of an atom, and
$r=(x^2+y^2)^{1/2}$ is the distance from the guide axis.

For a given transverse confinement, the beam is completely
characterized by its flux $\Phi$, its temperature $T$ and its mean
longitudinal velocity $v$. The flux is on the order of $5\times
10^9$ atoms/s. The temperature of the guided beam is deduced from
a radio-frequency (rf) spectroscopy technique \cite{prlrb2}. This
technique allows for the determination of the temperature with an
accuracy of typically 3\%~\cite{lahaye05}. For our parameters, we
find $T=640\;\mu$K. With this flux, temperature and mean
longitudinal velocity, the beam is clearly in the collisional
regime \cite{prlrb2}.

To demonstrate the implementation of evaporation by using a
ceramic surface, we have to remove atoms from the beam selectively
in position. This is accomplished by the application of a
transverse magnetic field $B_\perp$ that shifts the minimum of the
transverse confinement towards the cylindrical edge of the ceramic
surface (see Fig. \ref{figure1}.b). This magnetic field is
generated by a pair of coils with a mean radius 45~mm and
separated by a distance of 100~mm. Such a two-coils configuration
is particularly well suited for our demonstration since it
essentially results in a field perpendicular to the guide axis
over the whole range of interest, the longitudinal component being
negligible around the guide axis.

\begin{figure}[t!]
\centering\begin{center} \mbox{\epsfxsize 3.4 in
\epsfbox{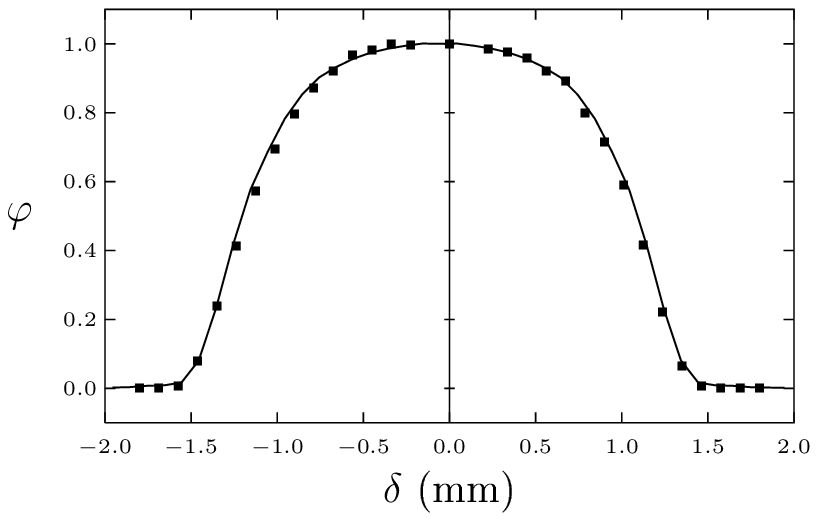}} \caption{Fraction of remaining atomic flux
$\varphi$ for a transverse shift of the position of the beam by
$\delta$ mm towards the edge of the ceramic surface: experimental
points ($\blacksquare$), and numerical simulation (solid
line).}\label{figure2}
\end{center}
\end{figure}

In presence of the coils field, the atoms are transversally
shifted by $\delta=B_\perp/b$. The confinement strength
essentially remains unaffected in the range of parameters where
the method has been used. We have checked that, in a zone without
a ceramic piece, the application of the transverse magnetic field
only results in a shift of the atomic beam. Indeed, even for
$\delta$ as large as 2~mm, the flux and temperature of the beam
remain constant within the accuracy of our measurements. This is
indeed the first prerequisite to validate our protocol.

Qualitatively, as the measured thermal transverse size of the beam
is at most of the order of the radius $R$, we expect that when the
line of minimum of energy has been shifted by the diameter of the
hole in the ceramic ($\delta \gtrsim 2R$), no atoms remain. For a
given displacement $\delta$, the fraction of remaining atoms
depends on the confining potential, the temperature, the radius
$R$ of the ceramic and the mean longitudinal velocity of the
atoms. Indeed, an atom with a low longitudinal velocity spends
more time in the evaporation zone and increases its probability of
being evaporated.

We have depicted on Fig. \ref{figure2} the fraction of remaining
atomic flux $\varphi$ as a function of the displacement $\delta$.
Even after a shift by $\delta=R$, no atoms remain. To understand
quantitatively the shape of this evaporation curve, we have
developed a Monte-Carlo numerical simulation. The initial
positions and velocities of the particles are randomly chosen
according to the thermal equilibrium distribution in the
semi-linear potential~(\ref{pot}). We use the experimental values
for the bias field $B_0$, the gradient $b$, the mean velocity $v$
of the beam and the temperature $T$. The motion of each atom is
governed by the Hamiltonian
$H=\boldsymbol{p}^2/2m+U(\boldsymbol{r})$, and calculated with a
symplectic fourth-order algorithm~\cite{symp}, interatomic
collisions are not taken into account. The surface of evaporation
is a cylinder of radius $R$ and length $\ell=5$ mm. To account for
the relative displacement between the atoms and the surface, the
cylinder is shifted off axis by a quantity $\delta$. Consequently,
the evaporation criterium reads $(x-\delta)^2+y^2 \ge R^2$. In the
numerical simulation, we assume a 100\% efficiency of evaporation.
This assumption is strongly supported by the excellent agreement
between the simulation (Fig.~\ref{figure2}, solid line) and the
experimental points (squares). The only adjustable parameter in
the simulated curve is a global shift on $\delta$, which reveals
that the ceramic piece is, in absence of transverse field,
out-of-center by approximatively 60~$\mu$m.

\begin{figure}[t!]
\centering\begin{center} \mbox{\epsfxsize 3. in
\epsfbox{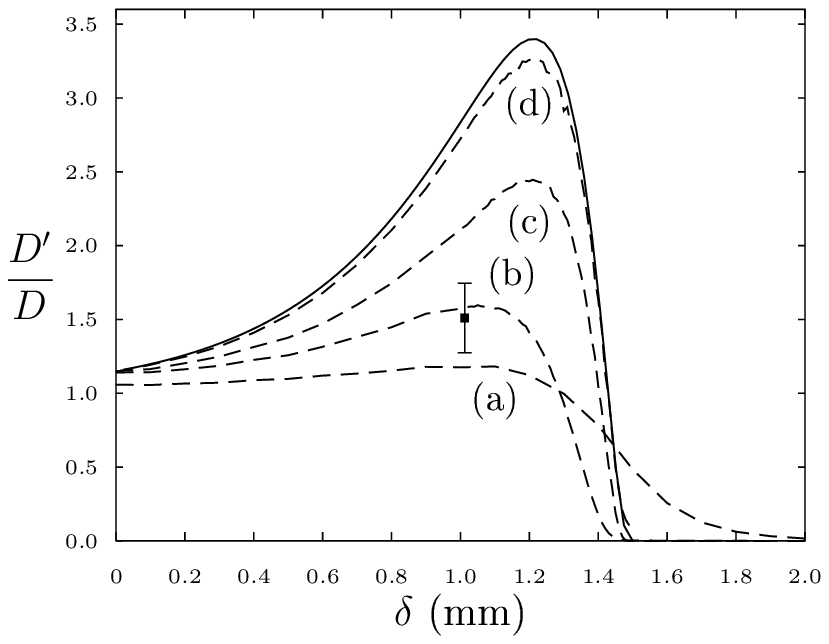}} \caption{Gain in phase space density resulting
from the evaporation on a ceramic surface after rethermalization.
Dashed lines: $D'/D$ deduced from the Monte-Carlo numerical
simulation for a surface length~$\ell=0.01$~mm (a), 5~mm (b),
10~mm (c) and 50~mm (d). Solid line: two-dimensional evaporation
where all atoms with a maximum distance from the center axis above
the radius $R-\delta$ are removed~\cite{epjd}. ${\blacksquare}$ is
the experimental point which corresponds to a gain in phase space
density of $1.5\pm0.2$. }\label{figure3}
\end{center}
\end{figure}

For a displacement $\delta=1$~mm and an upstream temperature of
640~$\mu$K, we have measured the temperature 3~m downstream the
ceramic piece. After such a distance the beam has completely
rethermalized to the lower temperature 500~$\mu$K. In combination
with the measured 40\% flux reduction  (Fig. \ref{figure2}), we
deduce a gain in phase space density of $1.5\pm0.2$. This result
is in good agreement with the prediction of our numerical
simulation (see Fig. \ref{figure3}) where one can infer the
downstream temperature and the gain in phase-space density by
computing the remaining flux and energy per atom after the
evaporation zone.

Unlike radio-frequency evaporation, this surface evaporation
occurs only along one dimension in position space. One may wonder
about the efficiency in terms of gain in phase space density and
collision rate of this one-dimensional evaporation. However, each
atom trajectory precesses around the center of symmetry axis since
the confining potential is not harmonic but semi-linear (see Fig.
\ref{figure1}.c). As a consequence, if an atom with adequate
transverse energy and angular momentum spends a long time with
respect to its oscillation period in the evaporation region, it
will be evaporated, irrespective of its initial position. In other
words, the real dimensionality of evaporation depends on the
length $\ell$ of the surface. We have depicted on Fig.
\ref{figure3}, the gain $D'/D$ in phase space density that we
deduce from our numerical simulation with our experimental
parameters for a length of the ceramic piece equal respectively to
0.01~mm (a), 5~mm (b), 10~mm (c) and 50~mm (d). The gain in phase
space density is clearly magnified if atoms spend a longer time in
the evaporation zone. Indeed if the surface is long enough, all
atomic trajectories reaching a distance from the minimum of the
magnetic potential larger than $R-\delta$ are evaporated. One then
realizes a true two-dimensional evaporation where all atoms above
a given radius are evaporated \cite{epjd} (solid line on Fig.
\ref{figure3}).

With a succession of ceramic pieces, it is possible in principle
to adjust in a very flexible way, using external magnetic fields,
the degree of evaporation by deflecting locally the atom
trajectories towards the surface. An alternative way would consist
in decreasing the radius of the successive pieces placed in the
ultra-high vacuum chamber which accommodates the magnetic guide.
This technique is particularly well suited for experiments aiming
at achieving a cw atom laser.

\begin{acknowledgements}
We are indebted to Jean Dalibard for a careful reading of the
manuscript. We acknowledge fruitful discussions with the ENS laser
cooling group, and financial support from the D\'el\'egation
G\'en\'erale pour l'Armement. Z.~W. acknowledges support from the
European Marie Curie Grant MIF1-CT-2004-509423. G.~R. acknowledges
support from the D\'el\'egation G\'en\'erale pour l'Armement.
\end{acknowledgements}


\begin{thebibliography}{99}

\bibitem[*]{byline}
Unit\'e de Recherche de l'Ecole Normale Sup\'erieure et de
l'Universit\'e Pierre et Marie Curie, associ\'ee au CNRS.

\bibitem{hydrogen}
C. Lovelace, C. Mehanian, T. J. Tommila, and D. M. Lee, Nature
(London) \textbf{318}, 30 (1985); H. F. Hess, G. P. Kochanski, J.
M. Doyle, N. Masuhara, D. Kleppner, and T. J. Greytak, Phys. Rev.
Lett. \textbf{59}, 672 (1987); N. Masuhara, J. M. Doyle, J. C.
Sandberg, D. Kleppner, and T. J. Greytak, Phys. Rev. Lett.
\textbf{61}, 935 (1988).

\bibitem{rmpnobelbec}
E. A. Cornell and C. E. Wieman Rev. Mod. Phys. \textbf{74}, 875
(2002); W. Ketterle, Rev. Mod. Phys. \textbf{74}, 1131 (2002).


\bibitem{Mandonnet00}
E. Mandonnet, A. Minguzzi, R. Dum, I. Carusotto, Y. Castin and J.
Dalibard, Eur. Phys J. D \textbf{10}, 9 (2000).

\bibitem{lahaye05}
T.~Lahaye, Z.~Wang, G.~Reinaudi, S.~P.~Rath, J.~Dalibard and
D.~Gu\'{e}ry-Odelin, Phys. Rev. A \textbf{72}, 033411 (2005).

\bibitem{prlrb2}
T. Lahaye, J. M. Vogels, K. J. G\"unter, Z. Wang, J. Dalibard, and
D. Gu\'ery-Odelin, Phys. Rev. Lett. {\bf 93}, 093003 (2004).

\bibitem{cornell}
D. M. Harber, J. M. McGuirk, J. M. Obretch, and E. A. Cornell, J.
Low Temp. Phys.  \textbf{133}, 229 (2003).


\bibitem{zeemanslower}
H. J. Metcalf and P. van der Straten, {\sl Laser Cooling and
Trapping}, (Springer-Verlag, New York, 1999).

\bibitem{launch} A. Clairon, C. Salomon, S. Guellati. and W.D. Phillips,
Europhys. Lett., {\bf 16}, 165 (1991).

\bibitem{symp}
H. Yoshida, Celest. Mech. Dyn. Astron. {\bf 56}, 27 (1993).

\bibitem{epjd}
T. Lahaye and D. Gu\'{e}ry-Odelin, Eur. Phys. J. D \textbf{33}, 67
(2005).

\end{thebibliography}
\end{document}